\documentclass[a4paper]{ifacconf}

\usepackage{graphicx,amsmath,url}      
\usepackage[round]{natbib}             

\usepackage{booktabs}
\usepackage[table,xcdraw]{xcolor}
\usepackage{lipsum}

\usepackage{placeins} 

\def\portugues{0} 

\ifx\portugues\undefined
\def\portugues{0}
\fi

\if\portugues0
   \usepackage[english]{babel}
  \else
   \usepackage[spanish,brazil,english]{babel}
\fi

\usepackage[T1]{fontenc}

\usepackage[utf8]{inputenc}

\usepackage{ae}

\if\portugues1
 { }
 { }
 { }
 
\fi

\begin{document}

\if\portugues1

%
\selectlanguage{brazil}
	
\begin{frontmatter}

\title{FED-HARGPT: A Hybrid Centralized-Federated Approach of a Transformer-based Architecture for Human Context Recognition} 

\thanks[footnoteinfo]{Reconhecimento do suporte financeiro deve vir nesta nota de rodapé.}

\author[First]{Primeiro A. Autor} 
\author[Second]{Segundo B. Autor} 
\author[Third]{Terceiro C. Autor}

\address[First]{Faculdade de Engenharia Elétrica, Universidade do Triângulo, MG, (e-mail: autor1@faceg@univt.br).}
\address[Second]{Faculdade de Engenharia de Controle \& Automação, Universidade do Futuro, RJ (e-mail: autor2@feca.unifutu.rj)}
\address[Third]{Electrical Engineering Department, 
   Seoul National University, Seoul, Korea, (e-mail: author3@snu.ac.kr)}

\selectlanguage{english}
\renewcommand{\abstractname}{{\bf Abstract:~}}
\begin{abstract}                

\vskip 1mm
\selectlanguage{brazil}
{\noindent \bf Resumo}:  As instruções abaixo são linhas gerais para a preparação de artigos para conferências e simpósios da Sociedade Brasileira de Automática (SBA) usando como base o estilo IFAC. Instruções de submissão podem ser encontradas no sistema de submissão de artigos ou no {\em website} do congresso.
\end{abstract}

\selectlanguage{english}


\selectlanguage{brazil}

\end{frontmatter}
\else
%

\begin{frontmatter}

\title{FED-HARGPT: A Hybrid Centralized-Federated Approach of a Transformer-based Architecture for Human Context Recognition} 

\thanks[footnoteinfo]{Sponsor and financial support acknowledgment
goes here. Paper titles should be written in uppercase and lowercase
letters, not all uppercase.}

\author[First]{Wandemberg Gibaut}
\author[First]{Alexandre Osorio}
\author[First]{Amparo Muñoz}
\author[First]{Sildolfo F. G. Neto}
\author[First]{Fabio Grassiotto}


\address[First]{Hub de Inteligência Artificial e Arquiteturas Cognitivas (H.IAAC)\\ Eldorado Research Institute, Campinas-SP, Brazil
  (e-mail: \{wandemberg.gibaut,alexandre.osorio, fabio.grassiotto\}@eldorado.org.br)
}

\renewcommand{\abstractname}{{\bf Abstract:~}}   
   
\begin{abstract}                
The study explores a hybrid centralized-federated approach for Human Activity Recognition (HAR) using a Transformer-based architecture. With the increasing ubiquity of edge devices, such as smartphones and wearables, a significant amount of private data from wearable and inertial sensors is generated, facilitating discreet monitoring of human activities, including resting, sleeping, and walking. This research focuses on deploying HAR technologies using mobile sensor data and leveraging Federated Learning within the Flower framework to evaluate the training of a federated model derived from a centralized baseline. The experimental results demonstrate the effectiveness of the proposed hybrid approach in improving the accuracy and robustness of HAR models while preserving data privacy in a non-IID data scenario. The federated learning setup demonstrated comparable performance to centralized models, highlighting the potential of federated learning to strike a balance between data privacy and model performance in real-world applications.
\end{abstract}


\end{frontmatter}
\fi

\section{Introduction}

With the increasing ubiquity of edge devices, such as smartphones and wearable technologies, there has been a notable increase in the generation of private data from wearable and inertial sensors, which enhance discreet monitoring of human activities, including resting, sleeping, walking, and stress level \cite{Annual2018}. Integrating private sensor data with advanced Artificial Intelligence techniques is gaining considerable interest in consumer products and industrial systems. This study focuses on implementing Human Activity Recognition (HAR) technologies and techniques using mobile sensor data (such as accelerometers and gyroscopes) and their application in edge devices. Additionally, this work explores the utilization of Federated Learning (FL) within the structure of the Flower framework to assess the training of a federated model derived from a centralized baseline. This evaluation is carried out through an experiment using the ExtraSensory dataset. This non-IID dataset aims to validate activity recognition in real-world conditions, thus moving closer to practical applications in everyday environments \citep{vaizman2017recognizing}.



The contributions of this paper are as follows.
\begin{itemize}
    \item Presenting a lightweight, Transformer-based model finetuned to an unrestricted HAR problem; 
    \item Presenting a hybrid centralized-federated approach to achieve a compromise between good performance, data privacy, and personalization, in a scenario where data is non-IID (Non-Independent and Identically Distributed);
    \item Comparative analysis on an unrestricted dataset, where data was collected without explicitly indicating to participants how to use the devices.
    
\end{itemize}



The remainder of the paper is organized as follows. Section \ref{sec:sec2} presents related works. Section \ref{sec:sec3} presents the background theory involved. Section \ref{sec:sec4} presents the methodology and experiments. Section \ref{sec:sec5} presents the results and the conclusion, and Section \ref{sec:secNew} presents a brief discussion.

\section{Related Works}
\label{sec:sec2}

Recently, significant progress has been made in the field of Human Activity Recognition, mainly using Multilayer Perceptrons (MLP) and other architectures to identify activities from mobile sensors. For example, the research by Mantyjarvi et al. used waist-worn accelerometers to identify a specific set of body movements, combining wavelet transform with principal component analysis and independent component analysis to generate features \cite{mantyjarvi2001recognizing}. An MLP classifier was used for classification, achieving recognition accuracies between 83-90\% for different human motions. Similarly, Kwapisz et al. used the built-in accelerometer of a smartphone, mounted on a front pant pocket, to recognize six body states, achieving up to 98\% accuracy in certain activities, although performance varied between activities \citep{kwapisz2011activity}. The model was trained on features derived from descriptive statistics and the timing of peak values in sinusoidal waves related to activities.

Additional advances in HAR have been demonstrated through the integration of sensors from smartphones and smartwatches. Some researchers explored the benefits of sensor fusion, showing significant improvements in detecting activities, including those associated with harmful habits such as smoking \citep{guiry2014multi,shoaib2015towards}. Guiry et al. evaluated five algorithms --- C4.5, CART, Naive Bayes, MLP, and Support Vector Machines --- reporting perfect accuracy across all instances. Shoaib et al. assessed three algorithms —Support Vector Machine, k-Nearest Neighbors, and Decision Tree — analyzing performance across individual sensors and their combinations.

The challenge of generalizing from controlled experimental conditions to real-life situations was highlighted by Kerr et al., who noted that data collected under controlled settings often performs poorly in natural environments (\cite{kerr2016objective}). Addressing similar concerns, Natarajan et al. explored the issues arising from using data collected in a laboratory to train classifiers that are then applied in field settings, such as discrepancies in class and sensor feature distributions and the difficulty in obtaining reliable ground-truth labels in uncontrolled environments \cite{natarajan2016domain}. Ermes et al. developed a system that enables participants to self-report activities via a personal digital assistant, selecting specific activities and contexts \citep{ermes2008detection}. In contrast, Choudhury et al. developed a system aimed to practical context recognition by being unobtrusive and fostering natural user behavior \citep{choudhury2008mobile}. Khan et al. provided a smartphone and collected data from participants in their natural environments for a month. 

Recent implementations of Federated Learning within Mobile Edge Computing (MEC) systems have also been explored to manage resources efficiently in distributed environments. Nishio and Yonetani integrated federated averaging in a practical MEC framework, allowing an operator to manage heterogeneous client resources effectively \cite{nishio2019client}. Wang et al. addressed the challenge of resource limitation in MEC systems by implementing various ML algorithms, including linear regression, SVM, and CNN, using federated averaging to optimize both computing and communication resources \citep{wang2019adaptive}. He et al. proposed FedGKT to mitigate computing limitations on edge devices, where each device trains only a portion of a full ResNet model to reduce computational overhead \citep{he2020group}.

Furthermore, \cite{vaizman2017recognizing} introduced the ExtraSensory dataset --- a rich, publicly available dataset for HAR collected in unconstrained environments --- and developed individual logistic regression classifiers to identify self-reported contextual data. They also designed a unified neural network model to tackle context identification as a multi-label classification problem, modifying the objective function to suit unconstrained data. This dataset and the corresponding application were presented as open-source resources, which facilitates further research and application development.  Besides the original paper, other works used the ExtraSensory as the main dataset, e.g., \cite{vaizman2018context} where the same research group presented a unified neural network model to solve a multi-label problem, \cite{fazli2020hhar}, who applied a hierarchical classification using a Deep Neural Network to categorize six primary labels, enhancing the precision in identifying activities like standing, running, or lying down. In addition, \cite{gibaut2022toward} presented a hybrid approach in which some clients are randomly selected to compose a base model, while others are used in Federated Learning rounds. Also, \cite{osorio2024transfer} presented a method using Federated Transfer Learning to reduce the computational cost of training the HAR model on smartphones for highly imbalanced datasets like ExtraSensory.

Other works utilize Federated Learning strategies for HAR problems. Sozinov et al. identified a trade-off between communication cost and the complexity of a model and proposed a method for erroneous client rejection \citep{sozinov2018human}. Xiao et al. presented a novel Federated Learning approach for HAR using wearable devices \citep{xiao2021federated}. Cheng et al. proposed a prototype-based aggregation method for Federated Learning in HAR, which can facilitate efficient communication among heterogeneous clients \citep{cheng2023protohar}.

Regarding Transformers-based models to HAR, Ji et al. use GPT-4 as a zero-shot learner for inference in structured data \citep{ji2024hargpt}. To the best of our knowledge, there is no previous work where a Transformers-based model was trained on an unbalanced, non-IID dataset for Human Activity Recognition.

\section{Theorical Background}
\label{sec:sec3}

This section presents the theoretical aspects involved in the present work. Each subsection discusses a specific topic.

\subsection{Human Activity Recognition}
Human Activity Recognition (HAR) refers to the process of identifying human actions or behaviors from observed data input, typically derived from various sensors or video recordings \citep{lara2012survey}. This multidisciplinary field intersects with areas such as computer vision, signal processing, machine learning, and ubiquitous computing, leveraging data-intensive approaches to discern patterns indicative of different physical activities. The core objective of HAR is to automatically detect and classify a wide range of human movements or activities, such as walking, running, sitting, or more complex sequences of movements, through the analysis of sensor-generated data or visual cues. 


In the domain of machine learning, HAR systems employ a variety of techniques ranging from traditional methods such as Decision Trees, Support Vector Machines (SVM), and Hidden Markov Models (HMM) to more contemporary deep learning approaches, including Convolutional Neural Networks (CNN) and Recurrent Neural Networks (RNN) \citep{straczkiewicz2021systematic}. These models are trained on labeled datasets to recognize patterns and features associated with specific activities. Deep learning, in particular, has shown remarkable success in HAR due to its ability to learn complex, hierarchical features from raw data, significantly increasing the accuracy of activity recognition \citep{gu2021survey}.


\subsection{Transformers}

Transformers are a groundbreaking class of models in natural language processing (NLP) and beyond, which have revolutionized the way machines understand and generate human language. Introduced in 2017 \citep{vaswani2017attention}, Transformers eschew the sequential processing paradigm of their predecessors, such as Recurrent Neural Networks (RNNs) \citep{bengio1994learning} and Long Short-Term Memory networks (LSTMs) \cite{hochreiter1997long}, in favor of a mechanism called self-attention. This innovation enables the model to weigh the importance of different words within a sentence, regardless of their positional distances from one another, thereby capturing the context more effectively and efficiently.


 The Transformer's ability to handle sequences of variable lengths and its proficiency in capturing long-range dependencies within the data makes it particularly powerful for tasks such as machine translation, text summarization, sentiment analysis, and, more recently, in models like GPT \citep{radford2018improving} for generative text tasks and BERT \citep{devlin2018bert} for understanding context and meaning in text. The adaptability and efficiency of this architecture have paved the way for their application in a wide range of tasks beyond NLP, including computer vision and generative tasks \citep{hudson2021generative}.

\subsection{Federated Learning}

Federated Learning is an innovative approach to machine learning (ML) that enables the training of models on multiple decentralized devices without the need to exchange local data \citep{mcmahan2017communication}. This paradigm shifts away from traditional ML methodologies, which typically require vast amounts of data in a single location. Developed to address growing concerns over privacy, data security, and sovereignty, Federated Learning allows the collaborative learning of a shared model while ensuring that sensitive data remain on the user's device, enhancing privacy and data protection.

At the heart of Federated Learning is bringing the model to the data, rather than the conventional approach of bringing the data to the model. In this setup, a global model is initially trained and distributed to all participants (e.g. devices or local servers). Each participant then trains the model on its local data to produce updated models or gradients aggregated on a central server or through a decentralized mechanism. This aggregated model is then improved iteratively, leveraging insights obtained from each participant's data without actually accessing it.


Federated Learning presents numerous advantages, particularly in fields where data privacy is paramount, such as healthcare, finance, and personal services \citep{yang2019federated}. For example, in healthcare, FL can enable the development of predictive models for disease diagnosis by learning from diverse datasets across multiple institutions without sharing patient data \citep{antunes2022federated}. This protects patient privacy and allows more robust and generalized models by learning from a wide range of demographic and geographic data.

However, Federated Learning also poses challenges, including managing communication overheads \citep{luping2019cmfl}, ensuring model convergence across diverse and potentially unbalanced datasets \citep{servetnyk2020unsupervised}, and safeguarding against malicious participants aiming to compromise the model or infer sensitive information. Despite these hurdles, the potential of Federated Learning to facilitate privacy-preserving, collaborative machine learning makes it a compelling area of research and development, promising significant advancements in how AI systems are trained and deployed in privacy-sensitive applications.

\subsection{The ExtraSensory dataset}

The ExtraSensory dataset includes sensor readings collected from smartphones and smartwatches used by 60 participants engaged in various physical and everyday activities across diverse locations \citep{vaizman2017recognizing}. The dataset captures information from a range of sensors, such as accelerometer, gyroscope, magnetometer, GPS, audio, location, and phone state indicators. Combining raw measurements, statistical summaries, and pseudo-sensor data, the dataset encompasses a total of 225 features.

\begin{figure}[!htb]
\centering
    \includegraphics[width=1.0\columnwidth]{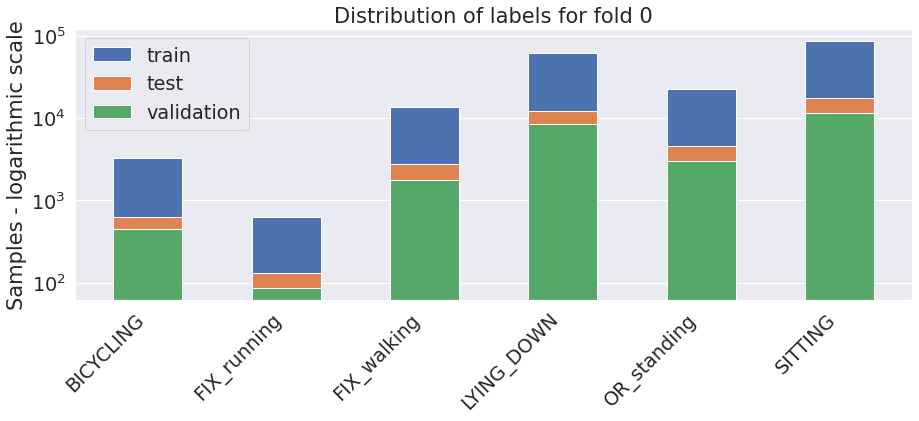}
\caption{Logarithmic-scale histogram for one of the cross-validation folds, showing the distribution of six of the labels of the ExtraSensory dataset, illustrates the significant imbalance in label distribution.}
\label{fig:extrasensory_data}
\end{figure}

Overall, the dataset comprises over 300,000 minutes of real-world recordings, where participants used the devices freely, rather than following predefined tasks in a controlled environment. As a result, the distribution of activity labels is significantly imbalanced, as shown in Figure \ref{fig:extrasensory_data}.

\section{Methodology and Experiments}
\label{sec:sec4}

The primary objective of this research is to explore the application of Federated Learning techniques to Human Activity Recognition using a Transformer-based model. This task is inherently complex due to the large peculiarity in human body, motion and phone use and demands a high degree of privacy. We sought to develop a model to recognize all 51 ExtraSensory labels. Some categories are mutually exclusive, like 'running' and 'lying down', but most can be simultaneously selected, so the classification problem is multi-label. The dataset was divided into training (including validation) and testing sets with an 80-20 split, maintaining a proportional representation of these activities to address the inherent class imbalance.

Similar to our previous work, cross-validation was employed to evaluate the model's generalization, wherein the dataset was distributed among various FL clients \citep{gibaut2022toward}. Specifically, the data from 60 subjects was segmented into five folds. Each fold comprised data from a distinct group of 12 subjects, designated as clients, while the aggregate data from the remaining 48 subjects were utilized to train a base model. This first training establishes a baseline from which the FL system fine-tunes the global model. The initial weights of the global model are set with the values derived from this centralized pre-training phase, which involves a comprehensive subset of 48 individuals distinct from those in each client-specific fold. This strategy is expected to decrease the number of training epochs required to adapt the model in a federated context, a critical consideration for subsequent deployment on edge computing devices with constrained processing power and memory, such as Android smartphones.

\subsection{FED-HARGPT}
In the present study, the model architecture employed is derived from the GPT-2 framework \citep{radford2019language}, albeit with some tailored modifications to suit the specific requirements of this research. The choice of GPT-2 as the base model is predicated on its relatively lightweight structure compared to more recent large language models (LLMs). This characteristic of GPT-2 facilitates the deployment of multiple model instances within a single traditional device --- AMD Ryzen Threadripper 3960X 24-Core Processor with two RTX 3090 GPUs, of which only one was used ---, enabling efficient Federated Learning processes across distributed computing environments. 


One of the adaptations made to the original GPT-2 architecture to better align with this work's objectives involved removing the embedding layer that typically serves as the entry point for input data in the standard GPT-2 model. In its place, a trainable linear layer was introduced, providing a more flexible mechanism for initial data processing that is adaptable to the nuances of the input data specific to this study, such as being continuous by nature and cannot be fully represented in a discrete space.

Further enhancements to the model include adding two distinct layers atop the base GPT-2 structure. The first addition is another linear layer, designed to refine the transformations applied by earlier layers and prepare the data for final classification tasks. Following this, an extra layer featuring a 'tanh' activation function was incorporated. The design of this layer includes a number of neurons that correspond directly to the number of labels in the multilabel classification problem addressed by this research, allowing for precise output mapping according to the defined categories.


Integrating these modifications into the GPT-2 architecture enables the tailored model to effectively handle the specific requirements of the multilabel classification task within a Federated Learning context. By adapting the model in this manner, it is better equipped to process and classify the diverse and distributed data typically encountered in Federated Learning scenarios, thereby enhancing the overall efficacy and applicability in practical settings.


\subsection{Centralized Training}

The initial training phase involved a usual, centralized Machine Learning process using aggregated data from 48 participants. The chosen model for this task was the already explained custom model, referred to in this work as FED-HARGPT (an acronym for Federated Human Activity Recognition GPT). A Random Search technique was employed to optimize the hyperparameters. We ran for 400 epochs, a batch size of 64, and the parameters adjusted during this process included:  

\begin{itemize}
    \item number of transformer layers: 1, 2, 3, 4, 6 or 12
    \item hidden size: 48, 96, 192, 384 or 768
    \item number of positions: 32, 64, 128 or 256
    \item learning rate: log uniform distribution ranging from 1e-5 to 1e-1
\end{itemize}

The primary metric targeted for optimization was \textit{Balanced Accuracy}, a suitable measure for datasets with significant class imbalance, as it avoids misleading performance assessments that can arise from underrepresented classes. Balanced Accuracy is calculated as follows:

$$ Balanced Accuracy = 0.5*(specificity + sensitivity)$$

where 
$$specificity = tn / (tn + fp) $$
$$sensitivity = tp / (tp + fn) $$

 and \textbf{tp}, \textbf{tn}, \textbf{fp} and \textbf{fn} stand for True Positive, True Negative, False Positive and False Negative, respectively.


The optimal model found by the hyperparameters optimization were: 
\begin{itemize}
    \item \textit{hidden\_size}: 384
    \item \textit{n\_positions}: 128
    \item \textit{transformers\_layers}: 4
    \item \textit{learning rate}: 1e-5
\end{itemize}

For each folder, we trained a model with the learning rate increased to \textit{4e-5} and the number of epochs to \textit{20000}. This number of epochs was chosen considering the grokking effect on model training \citep{power2022grokking}.


\subsection{Federated Training}

The clients were configured using the Flower Federated Learning framework \cite{beutel2020flower}, initiating with five models that had previously undergone training. This setup involved individual threads on a Linux system, as previously mentioned. Each client independently executed training and testing rounds using their local datasets in this federated system. After each round, the clients communicated their model weights to the FL server, which operates locally on the same system. 


The Flower framework offers flexibility to adjust various parameters of the Federated Learning strategy. In this particular setup, the FedAvg Strategy \citep{mcmahan2017communication} was employed with specific configurations to optimize the learning process. We set both fit and evaluation fractions to $1.0$, minimum available clients to 12 (all of them), batch size of 64, local epochs to 2000, and number of rounds to 4. 


The Federated Learning protocol was configured to execute over five distinct phases, consistent with the cross-validation scheme previously described. This configuration utilized twelve threads, each hosting an instance of a Flower client. During each round of cross-validation, clients initiated the process by loading the base model and then proceeded to train using their designated local datasets. This method facilitated a thorough evaluation of the model performance, using varied data subsets to rigorously assess its generalization and efficacy in the context of Federated Learning.


\section{Results and Conclusions}
\label{sec:sec5}



The first conclusion we found in our analysis was that, consistent with our previous work \citep{gibaut2022toward}, there is a low correlation between the number of samples available and the balanced accuracy of each client. Also, as seen in Table \ref{table:per_fold}, there was a considerable difference in metrics between the best and the worst fold (around 7.8\% in Balanced Accuracy), lying between 0.718 and 0.779.

\begin{table}[!htb]
\centering
\caption{Folds' statistics outlining  differences in BA between the best and the worst fold}
\label{table:per_fold}
\setlength{\tabcolsep}{12pt}
\begin{tabular}{@{}crrr@{}}
\toprule
fold                & mean balanced accuracy (BA) \\ \midrule
0                   & \multicolumn{1}{c}{0.754}           \\
1                   & \multicolumn{1}{c}{0.728}           \\
2                   & \multicolumn{1}{c}{0.718}           \\
3                   & \multicolumn{1}{c}{0.759}           \\
4                   & \multicolumn{1}{c}{0.779}           \\ \bottomrule
\end{tabular}
\end{table}

\begin{figure}[!htb]
\centering
    \includegraphics[width=1.0\columnwidth]{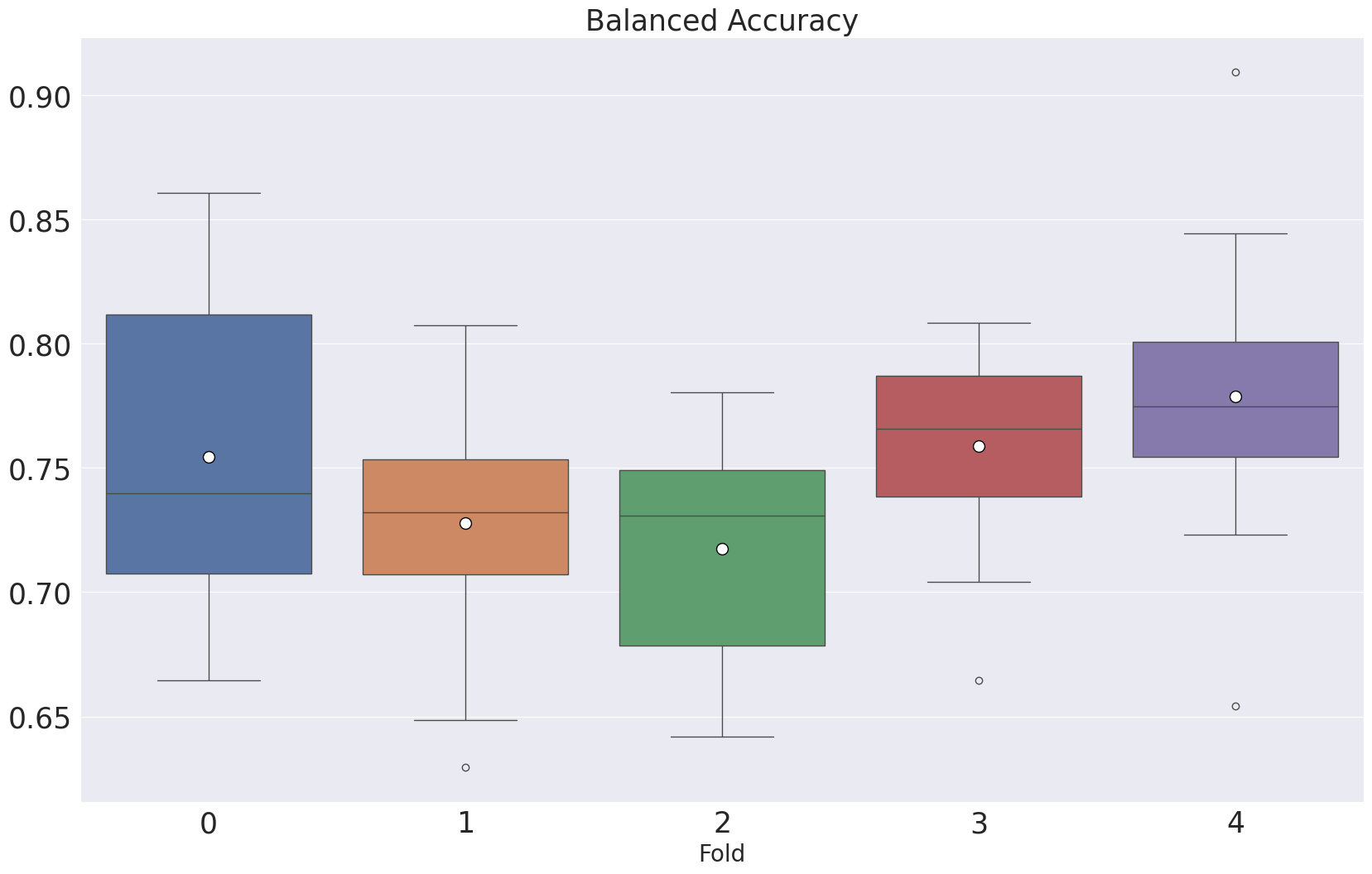}
\caption{Box plot for each fold in the Federated Learning process. Note that the dispersion of the clients' BAs has a relatively stable median, but the size of the quartiles and tails of the distributions may present very different values. }
\label{fig:box_ba}
\end{figure}

Figure \ref{fig:box_ba} shows each fold's boxplot of the Balanced Accuracy. The distribution shows that, despite most of the clients' BA being between 0.71 and 0.82 (the first and third quartile of all folds are in this range), the individual result can be as low as 0.63 or as high as 0.90. We argue that this phenomenon is closely related both with the diversity of collected data and the nature of some activities: clients that have a higher diversity of labels will produce data better suitable for the classification task, while some activities are inherently challenging to infer from mobile sensor data.

\begin{figure}[!htb]
\centering
	\includegraphics[width=1.0\columnwidth]{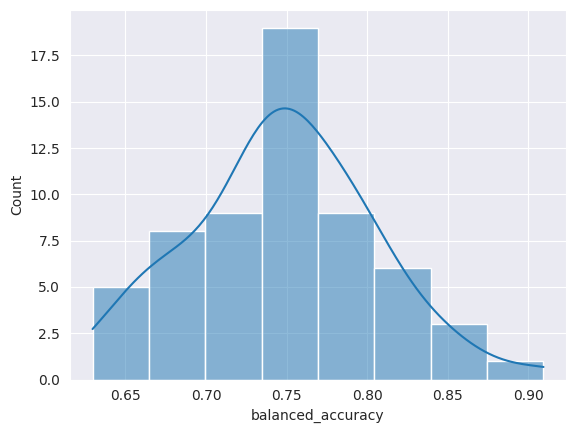}
\caption{Histogram of Balanced Accuracy results considering all folds. The blue curve is a Kernel Density Estimate (KDE) that helps visualize the data distribution. Most clients' BAs lie between 0.7 and 0.8, indicating some consistency}
\label{fig:hist}
\end{figure}

Figure \ref{fig:hist} shows the distribution of the Balanced Accuracy considering all five folds. The shape resembles a Gaussian distribution, with most of the curve being better than the established state-of-the-art (SotA) for the dataset being used. Again, we argue that discrepancies between the clients are, as mentioned before, due to the diversity of data between them and the nature of each activity.

Table \ref{table:ba} compares the metrics of the present work and those found in publications using the same dataset for all 51 labels \citep{vaizman2018context}. Numbers in parenthesis represent the number of neurons in hidden layers; that is, (8) would represent one hidden layer with eight neurons, while (16, 16) represents two hidden layers with 16 neurons each. The models of the present work achieved a mean performance slightly below SotA, but almost a third of clients (19 of 60) had better or equal performance than what was established, and the best client achieved outstanding results.

The experiments show that using a Transformer-based architecture, fine-tuning some data in a centralized way, and then further fine-tuning with Federated Learning can lead to good results for HAR problems. This approach leverages the benefits of pre-trained language models while maintaining data privacy and personalization. As a possible future work, Federated Learning strategies other than FedAvg could be tested, such as FedPer, which performs better in non-IID data scenarios \citep{arivazhagan2019federated}.

\begin{table}[!htb]
\centering
\caption{Comparison between our approach and state-of-the-art}
\label{table:ba}
\begin{tabular}{@{}ll@{}}
\toprule
Classifier                               & Balanced Accuracy \\ \midrule
MLP(8) \cite{vaizman2018context}        & 0.772 \\
MLP(16,16) \cite{vaizman2018context}    & 0.773 \\
\textbf{FED-HARGPT (mean)}                   & 0.747    \\
\textbf{FED-HARGPT (best folder mean)}       & \textbf{0.779}    \\
\textbf{FED-HARGPT (best client)}            & \textbf{0.909}    \\  \bottomrule
\end{tabular}
\end{table}

\section{Discussion}
\label{sec:secNew}

The experimental results from our study highlight the robustness and efficacy of the hybrid centralized-federated approach in Human Activity Recognition using a Transformer-based architecture. The Balanced Accuracy metrics demonstrate consistent performance across various data folds, with most accuracy values clustering around 0.75. This consistency affirms the reliability of our model in diverse scenarios and underscores the potential of Federated Learning to maintain high performance while ensuring data privacy. This capability is exciting in healthcare and finance, among others, where privacy concerns are high.

Our study opens paths for future work that addresses the scalability and efficiency of Federated Learning processes using Language Models. Techniques such as hierarchical Federated Learning and differential privacy can further optimize communication overhead and ensure robust model convergence, even with increasing data and device volumes. Expanding the scope of HAR by incorporating diverse sensor data and advanced machine learning techniques could yield better performance and broader applicability. The promising results from our study pave the way for leveraging hybrid centralized-federated approaches in various domains, such as smart cities and personalized health monitoring, ensuring data privacy while providing accurate, context-aware insights. This research sets a strong foundation for future innovations and applications in Federated Learning and AI-driven systems. Also, the use of a small LLM indicates the potential of such models to show good performance on complex problems like HAR on non-IID data.

In conclusion, our study demonstrates that a hybrid centralized-federated approach using a Transformer-based architecture is feasible and highly effective for Human Activity Recognition in non-IID data scenarios. The balanced performance across different data folds, combined with data privacy and security advantages, highlights the potential of Federated Learning in practical applications. As we move forward, continuous improvements and adaptations of this approach will be essential to address challenges like data diversity, better model personalization, and the full harnessing of its benefits across diverse domains. 


\section*{Acknowledgements}
The authors would like to thank the support of the Hub for Artificial Intelligence and Cognitive Architectures, a project founded by PPI-Softex/MCTI by grant 01245.003479/2024-10 through the Brazilian Federal Government.

\bibliography{ifacconf}             
                                                   







\end{document}